\documentclass[english,reqno,dvips,12pt]{article}

\usepackage{epsfig,graphicx,amsmath,amssymb,amsfonts,verbatim,amsthm,amscd}
\newtheorem{theorem}{Theorem}[section]
\newtheorem{lemma}{Lemma}

\theoremstyle{definition}
\newtheorem{definition}{Definition}
\newtheorem{example}{Example}

\theoremstyle{remark}
\newtheorem{remark}{Remark}

\begin{document}

\title{\textbf{Threshold levels in Economics}}
\author{\textbf{V.~P.~Maslov}}

\date{}

\maketitle

\begin{abstract}

In this paper, we present theorems specifying the critical values
for series associated with debts arranged in the order of their
duration.

\end{abstract}

\section{Introduction}

The mathematical theory used by economists up to now is
probability theory and optimization theory. These theories cannot
explain significant changes brought about by computerization just
as, at the beginning of the 20th century, the classical physics
could not interpret new experiments.

Nowadays the mathematical background of economics is the so-called
complexity theory (A.~N.~Kolmogorov, 1956) and the closely-related
theory of mathematical prediction (R.Solomonoff, 1964) as well as
some other arithmetic (addition rules) introduced by the author
\cite{Quadrat:x300}.

The threshold values guaranteeing economic security, which are
cited by the economists, are rather nebulous
(see for example~\cite{Dvorankov:x300}).

These values correspond to the violation of economic security,
while it is of interest to find the threshold values beforehand
so that correction measures can be taken.

Let me give a simple example of a critical number. Recall the
``trick'' of Korovjev, aka Fagot, in the novel ``Master and
Margarita'' by Mikhail Bulgakov. Banknotes of 10-ruble
denomination were scattered among the spectators at a variety
show. We assume that all the variants of permutation of $n$
banknotes among $k$ spectators are equiprobable (the chaos
assumption). In that case, if the number of spectators is greater
than $\sqrt{n} ln n$, then there is a large probability that
$k-\sqrt{n} ln n$ spectators will not get anything at all.

But if the spectators are combined into groups so that the number
of groups is $\sqrt{n}$, then there is a large probability that
all the groups will get some banknotes, which will be than divided
in a friendly manner among all the members according to the rules
of ordinary arithmetic. This represents the amount of
intervention of ordinary arithmetic (state management) into the
arithmetic of a chaotic market, which is determined by the excess
of indicators over the corresponding critical number.

In real economics, corporate bodies (firms), not natural persons,
are analogs of the spectators. Laws of ordinary arithmetic and
planning act within each firm where there is no economic
competition. But there is economic competition among firms.
 State regulatory intervention is necessary in cases where the critical number
is exceeded and some firms go bankrupt.

In the case of the bankruptcy of a firm, its employees become unemployed.
And, in that case, it is up to the state to take care of them,
otherwise, there may be trouble for the employed citizens.
Therefore, the state must intervene in free economic relations
and plan the corresponding actions where necessary. The amount
of such intervention depends on the excess of the predicted value
over the critical number.

As I have already pointed out on numerous occasions, another way
to fight the ``inexorable law of numbers'' is to introduce a
multi-currency system, which increases the barter component
\footnote{It means a combination of  the quantum statistics  of
identical particles (Bose statistics)  and  the classical
Boltzmann statistics.}.

Let me give an example.

Suppose we want to deposit two kopecks in two different banks
(see \cite{QuantEconomics:x300}; then we can say that there are
three possibilities: (1) put both kopecks into one of the banks;
(2) put both kopecks into the other bank; (3) one kopeck in one
bank, the other, in the other one. Here it is of no consequence
which of the two coins we deposit in the first bank and what is
its year of issue. Now imagine a situation in which we are
depositing one kopeck and one pence instead of two kopecks. In
that case, we have four options rather than three, because it is
significant which coin we placed in what bank, and so the variant
in which the coins are placed in different banks yields two
different options: (a) one kopeck to bank~1 and one pence to bank~2;
(b) one kopeck to bank~2 and one pence to bank~1.

The first case corresponds to the Bose--Einstein statistics that
leads, as is well known, to the Bose--Einstein condensate. In
financial mathematics, this phenomenon~\cite{ArXiv_2003:x300}
yields threshold values.  In the second case, the number of
variants corresponds to the Boltzmann statistics, which, however,
does not encompass such a phenomenon.

Let me give a second example. The members of a family wish to
change their flat into two flats. Nowadays, in Russia, the flat
in question is sold and then two flats are bought with the help
of real-estate agencies. This procedure takes at least two
months. In Soviet times, there were long exchange chains, which
can now be optimized using computers and data bases. If the price
of flats rapidly changes during the flat-exchange process, then,
due to the uniform variation of prices, the parties do not lose
anything if all the flats are exchanged at the same time. Since
such an exchange chain corresponds to  the Boltzmann statistics,
there is no Bose condensate.

There is yet another method, which was described in Bulgakov's novel:
the 10-ruble banknotes turned into sweet wrappers. But, in that case,
there is no more trust in 10-ruble banknotes, which leads to
a downturn in the economy, as is happening now in our country
where the banks are afraid of giving out ruble credits.

If the government does not intervene, then such processes may
occur spontaneously, leading to undesirable losses. There is no
way to avoid mathematical laws. As Admiral Kolchak\footnote{My
grandfather, Academician Pyotr Maskov, was a member of the Omsk
government at the time when Kolchak was War Minister and liked to
quote Kolchak's aphorism. Later, in 1919, Kolchak became known as
``The Supreme Ruler of Russia.''} put it, there is ``this
inexorable law of numbers'' by which he apparently meant
superiority in numbers.

The separation of the money supply from the debt total is an
approximation just as the quantum statistical physics of bosons
is an approximation of scalar quantum field theory in its
Hamiltonian version\cite{Masl_Shv_1:x300, Masl_Shv_2:x300,
Masl_Shv_3:x300}. In the latter theory, particles and holes can be
created and annihilated in the same way as money annihilates
debts.

There is also a rapid turnover in debts; they are annihilated and
created anew. However, in the mean, both debts and assets are
quite obvious even more so than money and money turnover.

If we use an electronic bank card for payments,
then this increases the rate of turnover in the same way
as e-mail increases the postal delivery rate.
However, we must find the weak limit with respect to these
rapid variations so as to obtain threshold values to be discussed
in what follows.

The concept of the mean rate of turnover defined in textbooks in
economics as the ratio of GDP to the amount of the available money
was meaningful only in the case of gold coins. But, nowadays, it
is the same weak limit, which is practically used when in
estimating bank assets.

In the same way, in the thermodynamic limit of the total system,
the number of particles $N^+$ and holes $N^-$ is preserved (the
model of ions--electrons in a plasma; see~\cite{Masl_2002:x300}).
Hence the sum
\begin{equation}\label{b0:x300}
N^+ + N^- =N_{\text{total}}
\end{equation}
is also preserved.

But if we consider a subsystem in which the number~$N^-$ has
increased due to external action, then, in the first equilibrium
approximation, it turns out that $N_{\text{total}}$ is preserved,
and hence $N^+$ is decreased.
 At the same time, the total energy after deduction of the infinite energy
of vacuum is preserved.
 Approximately,
\begin{equation}\label{b1:x300}
\varepsilon\sum_{-k}^\infty iN_i=\cal{E},
\end{equation}
where $\cal{E}$ is the total energy, $\varepsilon$ is the energy
of one particle, and
\begin{equation}\label{b2:x300}
\sum_{i=-k}^\infty N_i=N.
\end{equation}
 Making the replacement
$i+k=j$, we obtain
\begin{equation}\label{b3:x300}
\varepsilon\sum_{j=0}^\infty (j-k)N_{j-k}= \cal{E}.
\end{equation}
 Taking~\eqref{b2:x300} into account, we can write
\begin{equation}
\label{b4:x300} \varepsilon\sum_{i=0}^\infty iN_i={\cal E}
+\frac{k(k+1)}2\, N\varepsilon.
\end{equation}

 Here the term
$k(k+1)/2$ tends to infinity as $k\to \infty$.
 This is the analog of the energy of vacuum, which must be deducted.
 A rigorous proof of passages to the limit with deduction of infinities
is difficult.
 An attempt to justify these passages mathematically
(such as the passage to nonrelativistic quantum field theory) was
made in~\cite{Masl_Shv_1:x300},~\cite{Masl_Shv_2:x300},
and~\cite{Masl_Shv_3:x300}.

 How to interpret the law of conservation of the sum
$N^++N^-$ (the sum of assets and liabilities) in economics?
 Although a significant role in economics is played by the psychological factor,
this law, nevertheless, corresponds to mathematical laws in the
mean (after averaging over subjects and time).

 Let us assume that a young family borrows some money for education,
mortgage, and a car.
 This stimulates the family to earn more money.
 If one-half of the increased income is used to repay the loan,
and the other half goes toward increasing the assets, then we have
$N^++N^-= \text{const}$.

 Debts and assets treated separately are, in fact,
the result of averaging over rapid transitions
``credits$\leftrightarrow$incomes.''

 In practice, both debts (liabilities) and money (assets)
are calculated, although, as pointed out above, they are averaged
over rapidly varying turnovers ``money$\leftrightarrow$debts.''

 Therefore, pre-crisis American banks can be associated with,
for example, the time series of their assets and they can be
classified according to the size of their assets.
 The journal ``Forbes'' classifies the richest people
by the mean size of their fortune.
 From a mathematical point of view, we are not interested
in the names of persons or banks, but in their numbers ordered in
the same way as in ``Forbes,'' beginning with the greatest
fortune, and also their variation over sufficiently large time
intervals.
 Only from these time series, can we
determine economic threshold values by using the theorems given
below in this paper.

 Let us begin by considering debts.
 In addition to other features of a debt, such as the percentage,
each debt~$s$ is characterized by its period of repayment
(pay-out period)~$l$.
 The ratio of the debt size to its pay-out period is the quantity
of immediate interest to the borrower, because it determines the
amount of extra effort per day needed to repay the debt in time.
 If there are debts
$s_i$, $i=1,2,\dots, k$, with pay-out periods~$l_i$, then the sum
of the ratios of debts to their pay-out periods represents the
amount of extra money to be earned each day.
 The ratio of
$\sum_{i=1}^k s_i/ l_i$ to the amount of all the debts
$\sum_{i=1}^k s_i$ is the mean period of repayment of all debts.

 Here we present a criterion and a certain critical constant
(threshold level).
 If the mean period of repayment of debts is
greater than this level, then the borrower is bankrupt.
 This
quantity is just as precise as the number of spectators at
Koroviev's performance; the important thing is that it depends on
quantities specified by the series $\{s_i\}$ and $\{s_i/l_i\}$.
 This looks like a mathematical mystery.
 However, this phenomenon is related to a remarkable physical
fact, namely, the existence of the Bose--Einstein condensate for a
Bose gas.
 In our case, this means that if this number is greater
than the critical number, then the debts turn out to be long-term
ones, which usually are mortgages.

 People did not believe in this physical phenomenon for a long time,
but, finally, got used to it.
 And there exist many experiments
that confirm this phenomenon.
 In deterministic physics,
it has no easy explanation.
 But there is an explanation in economics.
 There are few experiments
involving a crisis: it seems that there is just one such
world-wide experiment.
 In this sense, economics can enrich physics, just as it helped the author
to explain the so-called $\lambda$-point in the Bose condensate
of Helium-4 and the Thiess--Landau model (see below).

 The war in Iraq required substantial loans, mostly internal ones
(such as from insurance companies) and a subsystem (financial
bodies and the defence industry) having received $N^-$ credits
(debts, holes) decreased the number of particles (the total
energy being constant) according to the law~\eqref{b0:x300}; this
means that the turnover of particles (the Bose-gas temperature)
increased or, as the economists put it, the subsystem has heated
up and there began an abrupt upturn in the economy of the
subsystem.
 Naturally, this resulted in an increase of mortgage (floating) interests,
 while most of the people living in mortgaged houses
 was not involved in the system where incomes were growing.
 Hence the mortgage crisis occurred and this was followed by the crisis
of insurance companies dealing with mortgages (and also long-term
credits), etc.
 This is a rather deterministic explanation of Bose condensate.
 It is up to physicists and, especially, specialists in mathematical physics
to try to carry over this explanation to scalar Hamiltonian field
theory and use it for the deduction of the Bose-condensate
phenomenon.

 Let us now turn to the opposite phenomenon, inflation
(it is opposite in the sense that, as shown above, the debt
crisis leads to deflation).

 Here the situation is more complex, but it easily explains
inflation at the time when the currency was in the form of gold
coins.
 One could also calculate the GDP at that time, but, in
contrast to the present time, one could calculate the number of
gold coins in circulation more precisely (although bills of
credit were also used, but they could hardly be used to buy goods
in a shop; they could be bought up in order to bring about a
bankruptcy).
 At
present, banknotes are seldom used in shops abroad, mostly bank
cards and cheques, which, in fact, are bills of credit.

 Therefore, in textbooks on economics, there is a tradition of
long standing to state the main law of economics as follows: the
mean rate of turnover of money (gold coins) is equal to the ratio
of the GDP to the total number of gold coins.

 The GDP of a country is composed of similar quantities associated
with its constituent parts (regions) and these quantities, in
turn, involve quantities corresponding to finer structures.
 In
such series, one can precisely calculate the threshold turnover
rate possessing the following property: if the mean turnover rate
is less than this threshold value, then inflation occurs, as a
result of which smaller coins (copecks in Russia) are no longer
used, while gold coins disappear into money boxes or chests.

\begin{remark}
 All the critical constants given above are highly sensitive
to a phenomenon called ``protectionism.''

 Suppose that, in our example, Koroviev took pity
on those spectators that did not get any banknotes and he gave
them one banknote in ``aid.'' Thus, he changed the situation by
equating such people to those who got one banknote by the
``market'' law.
 This is the same as if the unemployment benefit
were raised to the level of the wages of some categories of
workers.
 Then these workers would find themselves to be on the same ``zero''
level of unemployed persons.

 If the debts of the debtors are restructured to a longer period,
then those who took credits (such as mortgage) for such long
periods are threatened.

 Then the kind Koroviev should have added some money
to those spectators who had grabbed just one banknote.
 Thus, if
credit owners suffer as a result of ``protectionism,'' they
should be the ones to get help in the first place.
\end{remark}

 Since the law of conservation of energy in the physics
of particles--holes--photons that are created and annihilated at
a great rate is valid in the mean, it follows that the number of
particles and holes is preserved in the mean, too.
 Despite a tremendous rate of debts--money transitions,
in modern computer economics, assets and liabilities are
preserved in the mean
in the weak limit, and we can consider the time series of these quantities.

\section{Koroviev's trick and partition theory
in number theory}

 Let us return to Koroviev's trick.

 Assume that there were
$n$ banknotes and $k$ persons in the audience, where
$$
\text{ $k>\frac 1\pi \sqrt{\frac 32}\sqrt n \ln n$. }
$$
 We certainly assume that all versions of distributing $n$ banknotes
among $k$ persons are equiprobable.\footnote{Naturally, if the
spectators are well behaved and do not grasp bank notes from
other seats.
 Therefore, all the decompositions of an integer into
the sum of $k \leq k_0$ terms are equiprobable.}
 Let us state a theorem of number theory.

 Let
$n$ be a positive integer.
 By a {\textit partition} of
$n$ we mean a way to represent a natural number $n$ as a sum of
natural numbers.
 Let
$p(n)$ be the total number of partitions of~$n$, where the order
of the summands is not taken into account, i.e., partitions that
differ only in the order of summands are assumed to be the same.
 The number
$p_k(n)$ of partitions of a positive integer $n$ into $k$
positive integer summands is one of the fundamental objects of
investigation in number theory.

 In a given partition, denote the number of summands (in the sum)
equal to 1 by $N_1$, the number of summands equal to 2 by $N_2$,
etc., and the number of summands equal to $i$ by $N_i$.
 Then
$\sum N_i=k$ is the number of summands, and the sum $\sum iN_i$ is
obviously equal to the partitioned positive integer.
 Thus, we have
\begin{equation}\label{5*:x300}
\sum_{i=1}^\infty iN_i=n, \qquad \sum_{i=1}^\infty N_i=k,
\end{equation}
where the $N_i$ are natural numbers not exceeding~$k$.

 These formulas can readily be verified for the above example.
 Here all the families
$\{N_i\}$ are equiprobable.

 The distribution for the parastatistics
$N_i\le k$ and $\sum N_i=k$ is determined from the relations
\begin{equation}
\sum_{i=1}^n\frac1{e^{b(i+\kappa)}-1} -\frac
k{e^{bk(i+\kappa)}-1}= k,\qquad \sum_{i=1}^n\bigg(\frac
i{e^{b(i+\kappa)}-1} -\frac{ik}{e^{bk(i+\kappa)}-1}\bigg)=n,
\label{9:x300}
\end{equation}
where $b>0$ and $\kappa>0$ are constants defined from
relations~\eqref{9:x300}, $n/k$ is sufficiently large, and the
numbers $n$ and $k$ are also large, and we can pass (by using the
Euler--Maclaurin summation formula) to the integrals (for the
estimates for this passage, see~\cite{33:v851:x300}),
\begin{align}
\int_0^\infty\bigg(\frac1{e^{b(x+\kappa)}-1}
-\frac k{e^{bk(x+\kappa)}-1}\bigg)\,dx&\cong k, \label{10:x300}\\
\int_0^\infty\bigg(\frac x{e^{b(x+\kappa)}-1}
-\frac{kx}{e^{bk(x+\kappa)}-1}\bigg)\,dx&\simeq n. \label{11:x300}
\end{align}

 It can be proved that
$\kappa=0$ gives the number $k_0$ with satisfactory accuracy.
 Hence,
$$
k_{0}=\int_0^\infty\bigg(\frac1{e^{bx}-1}
-\frac{k_0}{e^{bk_{0}x}-1}\bigg)\,dx.
$$

 Consider the value of the integral (with the same integrand)
taken from $\varepsilon$ to $\infty$ and then pass to the limit as
$\varepsilon\to0$.
 After making the change
$bx=\xi$ in the first term and $bk_0x=\xi$ in the second term, we
obtain
\begin{align}
k_0&=\frac1b\int_{\varepsilon b}^\infty\frac{\,d\xi}{e^\xi-1}
-\int^\infty_{\varepsilon bk_0}\frac{\,d\xi}{e^\xi-1}=\frac1b
\int^{\varepsilon bk_0}_{\varepsilon b}\frac{\,d\xi}{e^\xi-1}\\
&\sim \frac1b\int^{\varepsilon bk_0}_{\varepsilon
b}\frac{\,d\xi}\xi =\frac1b\{\ln(\varepsilon
bk_0)-\ln(\varepsilon b)\}=\frac 1b\ln k_0. \label{13:x300}
\end{align}

 On the other hand, making the change
$bx=\xi$ in \eqref{11:x300}, we obtain
$$
\frac1{b^2}\int^\infty_0\frac{\xi \,d\xi}{e^\xi-1}\cong n.
$$
 This gives
\begin{equation}
b=\bigg({\sqrt n}\bigg/{\sqrt{\int_0^\infty\frac{\xi
\,d\xi}{e^\xi-1}}}\,\bigg)^{-1}, \qquad k_0=\frac12\frac{\sqrt
n}{\sqrt{\pi^2/6}}\ln n(1+o(1)).\label{15:x300}
\end{equation}

 Now let us find the next term of the asymptotics by setting
$$
k_0=c^{-1}n^{1/2}\ln c^{-1}n^{1/2}+\alpha n^{1/2}+o(n^{1/2}),
\qquad\text{where}\quad c=\frac{2\pi}{\sqrt6}\,.
$$
Furthermore, using the formula
$$
k_0=c^{-1}n^{1/2}\ln k_0
$$
and expanding $\ln k_0$ in
$$
\text{ $\frac\alpha{c^{-1}\ln c^{-1}n^{1/2}}\,,$ }
$$
we obtain
$$
\text{ $\alpha=-2\ln\frac c2$. }
$$
 Thus, we have obtained
the Erd\H os formula~\cite{Erdos:x300}.

 Let us show that, if
$k>k_0$, then $\kappa <\nobreak 0$.\nopagebreak

\begin{lemma}\label{lemma1:x300}
 Let
$n>k\gg k_0$.
 Suppose that
$\kappa=-\mu$ and $\mu>0$.
 In this
case, equations \eqref{10:x300} and \eqref{11:x300} have
solutions with $\mu
> k^{-1/2-\delta}$,
where $\delta >\nobreak 0$ is as small as desired.
\end{lemma}

\begin{proof}
 Indeed, consider relations \eqref{10:x300} and \eqref{11:x300}.
 Making the change
$\xi-\mu=\eta$, and then $b\eta=\varphi$,
 we obtain
\begin{align} n&=\frac{1}{b^2}\int_0^\mu \frac{k\xi
\,d\xi}{1-e^{-k\xi}}- \frac{1}{b^2}\int_0^\mu \frac{\xi
\,d\xi}{1-e^{-\xi}}+
\frac{1}{b^2}\int_0^\infty\bigg(\frac{\xi}{e^\xi-1}-\frac{k\xi}{e^{k\xi}-1}\bigg)
\,d\xi,\label{17:x300} \\
k&=\frac 1b\int_\varepsilon^\mu
\bigg(\frac{k}{1-e^{-k\xi}}-\frac{1}{1-e^{-\xi}}\bigg)\,d\xi
+\frac 1b\int_\varepsilon^\infty\bigg(\frac{1}{e^\xi-1} -
\frac{k}{e^{\xi k}-1}\bigg)\,d\xi.\label{18:x300}
\end{align}

 After making the change
$k\xi=x$ in the corresponding integrals, we see that
\begin{align}
 n&=\frac1{b^2}\frac1k\int_0^{\mu k}\frac{x
\,dx}{1-e^{-x}}+o\bigg(\frac1{b^2}\bigg),\label{19:x300}
\\*
k&=\frac1b \int_0^{\mu k}\frac{\,dx}{1-e^{-x}}+o\bigg(\frac{\ln
k}b\bigg) \label{20:x300}
\end{align}
as $k\to\infty$.
 Therefore,
$\mu k \ll \sqrt{k}$; for instance, $\mu k \geq
\sqrt{k^{1-\delta}}$ for any $1>\delta>0$.

 This relation can be satisfied provided that
$\mu >(\sqrt{k^{1+\delta}})^{-1}$.
 This proves the lemma.
\end{proof}

 The logarithm of the number of variants is the Hartley entropy, and
the entropy for a parastatistic is known.
\begin{equation}\label{f1:x300}
 S(k_0) \sim bn+ \int_0^\infty \ln\frac{1-e^{-(k+1)bx}}{1-e^{-bx}}
\,dx.
\end{equation}

 But if
$\kappa <0$, the entropy is of the form
\begin{equation}\label{f2:x300}
 S(k)=bn-\kappa k+\sum_{j=1}^\infty \ln
\frac{1-e^{-(k+1)b(j-\kappa)}}{1-e^{-b(j-\kappa)}}\sim bn-\kappa
k+\int_0^\infty
\ln\frac{1-e^{-(k+1)b(x-\kappa)}}{1-e^{-b(x-\kappa)}} \,dx.
\end{equation}

\begin{lemma}\label{lemma2:x300}
 The ratio
\begin{equation}\label{f3:x300}
\frac {p_k(n)}{p_{k_0}(n)}= e^{S(k)-S(k_0)} \geq e^{-(\mu
k)^{1-\delta}},
\end{equation}
where $\delta >0$.
\end{lemma}

\begin{proof}
 We can write
\begin{align}
\label{f4:x300} \nonumber
 S(k)&= bn - \mu k+\int_{-\mu}^0
\ln\frac{1-e^{-(k+1)b\xi}}{1-e^{-b\xi}}\,d\xi +\int_0^\infty
\ln\frac{1-e^{-(k+1)bx}}{1-e^{-bx}} \,dx
\\
&= S(k_0)-\mu k+\int_0^\mu \ln\frac{e^{(k+1)b\xi}-1}{e^{b\xi}-1}
\,d\xi.
\end{align}
 The estimate
$$
\int_0^\mu \ln\frac{e^{(k+1)b\xi}-1}{e^{b\xi}-1} \,d\xi
$$
yields the assertion of Lemma~2 after the calculation of the
maximum of the expression under the sign of logarithm.
\end{proof}

 A rigorous proof of the weak convergence to the Bose--Einstein
distribution of sequences corresponding to an ideal Bose gas of
dimension greater than two is given in a remarkable work of
A.~Vershik~\cite{Vershik:x300}.
 For the dimension
$d\leq 2$, this proof is valid for the parastatistic discussed
above.
 The
following theorem holds.

\begin{theorem}\label{theorem1:x300}
{1)} Suppose that $p_k(n)$ is the number of partitions of an
integer~$n$ into $k\leq k_0$ summands and $N_i$ is the number of
occurrences of the integer $i \leq k$ in these partitions.
 Then the difference
\begin{equation}\label{21:x300}
b \sum_{i=1}^k N_i \varphi(bi)-\int_0^\infty \varphi(x)\left\{
\frac{1}{e^{x+\kappa}-1} -\frac{k}{e^{k(x+\kappa)}-1}\right\}
\,dx,
\end{equation}
where the function $\varphi(x)$ is any bounded piecewise smooth
(with finitely many discontinuities) function continuous at the
points~$bi$, tends in probability\footnote{A random
variable~$X_n$ is said to tend to~$X$ \textit{in probability}, or
$X_n\xrightarrow{\mathsf P} X$, if, for any $\varepsilon>0$,\,
$\lim_{n\to\infty}{\mathsf P}(|X_n-X|>\varepsilon)=0.$ } to zero
as $k\to\infty$.

{2)} Let $k>k_0+o(k_0)$, and suppose that $k_0$ spectators got at
least one bank note each; hence the distribution~\eqref{21:x300}
is valid for them.
 Let
$N_0$ be the number of spectators without bank notes.
 In that case, obviously,
\begin{equation}\label{22:x300}
P \left(|N_0-(k-k_0)| > \delta_1k_0\right) \leq
e^{-(k-k_0)^{1/2-\delta}},
\end{equation}
where $\delta > 0$ and  $\delta_1 > 0$  are arbitrarily small and
do not depend on~$n$. Here $P$ is the ratio of the number of
variants satisfying the condition in parentheses in
\eqref{22:x300} to~$p_{k_0}(n)$.

Further, the occurrence of the numbers $i\leq A/b$, where $A$ is
an arbitrarily large number independent of~$b$, satisfies the
relation
$$
b \sum_{i=1}^{[A/b]} N_i \varphi(bi)-\int_0^\infty
\varphi(x)\left\{ \frac{1}{e^{x}-1}
-\frac{k_0}{e^{k_0x}-1}\right\} \,dx \xrightarrow{\mathsf P} 0
\qquad\text{as}\quad n\to\infty.
$$
\end{theorem}

\begin{proof}
 The formula for the parastatistic
$$
R(x)= \int_0^x\bigg(\frac1{e^{b(x+\kappa)}-1} -\frac
k{e^{bk(x+\kappa)}-1}\bigg)\,dx,
$$
given in physics textbooks obviously follows from the rigorous
relation for the parastatistic and Stirling's decomposition
formula.
 Sharper estimates~\cite{Masl_Vyugin:x300} are not needed for
the proof.

 Our further proof is quite similar to that of Vershik's theorem~\cite{Vershik:x300}.

 The use of the Euler--Maclaurin theorem and estimates for the
passage from sums to integrals yields Theorem~1.

 For
$k\leq\sqrt{n}$, using the Banach--Steinhaus theorem, we can
obtain weak convergence in formula~\eqref{21:x300}.

 The proof of inequality~\eqref{22:x300} follows from Lemmas~\ref{lemma1:x300}
and \ref{lemma2:x300}, because, by performing the change
$N_0=k-(k_0+o(k_0))$, we do not change the number of
partitions~${p_{k_0}(n)}$.
\end{proof}

\begin{remark}
 In physics, the parameter~$b$
is called inverse temperature and the parameter~$\kappa$ is the
chemical potential taken with a negative sign; the dimension~$d$
of the space in which a quantum ideal gas is considered is two in
our case.

 With regard to the value of the dimension, we shall follow the
physics literature in contrast to~\cite{QuantEconomics:x300},
where dimension was defined as the quantity half as large as the
physical dimension.
 The quantity~$d$
can be associated with the so-called Pareto distribution and its
rate of decrease.
 It is related to spectral density.

 In the physics literature, it is universally stated
(in contrast to~\cite{Temperley:x300}) that there is no Bose
condensate in the two-dimensional case (for
$d=2$).\footnote{Although it was stated in the old papers
of~\cite{Temperley:x300} and~\cite{Frish:x300} that there is a
condensate in this case.} The same statement is contained
in~\cite{Vershik:x300}.
 Indeed, formally, it is not a Bose
condensate, but the condensate of a parastatistic.
\end{remark}

 Let us consider a financial example in which the dimension
$d=2$ occurs.

\begin{example} \textsl{Japanese candles on the stock exchange.}
 On the stock exchange, the data on prices in a given time scale
(such as hour intervals) is given as four prices: opening price,
highest price, lowest price, and closing price.
 This allows
reduction in the amount of stored data as well in the time of
their processing.

 These four prices are usually represented as ``Japanese candles.''
 The extreme points of the upper and lower segments denote,
respectively, the highest and the lowest price for a given day,
while the upper and the lower base of the rectangle (body of the
candle) denote, respectively, the opening and closing prices if
the rectangle is white and the closing and opening prices if the
rectangle is black (see Fig.~6.3/1 in~\cite{QuantEconomics:x300})

 Thus, the Japanese candle is the symbol indicating the variation
of the price of a particular financial tool (share) on the stock
exchange. We shall number them in decreasing order of frequency.

 Not only Japanese candles on the stock market, but also the whole
world financial system relies on the positive values of~$\kappa$
($\kappa>0$), provided that $\kappa \to 0$ as $b\to 0$.
 Then it
demonstrates a sufficiently rapid growth.
 However, no wars must
break out; for $\kappa<\nobreak 0$, they may lead to the global
financial catastrophe.

 Lately, there is a growing tendency to group shares into
equivalent sets (similar to grouping words into descriptors).
 For
an investor, shares of one group may replace one another just as
pronouns and rare synonyms may replace repeating words and thus
change the statistics describing the occurrence of words.
 Investors try to avoid repetition to a greater degree than do
writers, who try to prevent the repetition of the same words: the
greater the range of goods bought, the less the risk to incur
losses.
 The rank on the graph~6.3/2 from~\cite{QuantEconomics:x300})
is the number of a candle (numbered in ascending order with
respect to candle size); in this case, the fractal
dimension~$\alpha$ is~$1$.
\end{example}

 Figuratively speaking, if the spectators in our example with the conjuring
trick are divided into groups so that the number of groups is
greater than the critical number, but exceeds it by only a small
amount, then the spectators in each group will be nervous about
the amount of bank notes they will be getting, and hence will be
active.
 But if it turns out that the number of groups is much
less than the critical number, than the spectators will be sure
that they will get a lot of bank notes with large probability,
and hence will be passive, or as P.~Milyukov put it, they will
``bask in the state of blissful idiotism'' or, which is still
worse, become obscurantists.

\section{The dimensions $d=1$ and $0<d<2$}

 Just as above, let us consider the dimension
$d=1$.
 If, in physics,
the dimension $d=2$ corresponds to a plane, then the dimension
$d=1$ corresponds to a line.

 First, consider the one-dimensional case of a Bose condensate, which
is important in physical problems, in the notation used in
statistical physics: $\cal E$ is the energy, $k=N$ is the number
of particles, $\varepsilon i$ are the energy levels, and $n=
{\cal E}/{\varepsilon}$.

 Let the dimension be
$d=1$.

 Define the constants~$b$
and~$\kappa$ from the following relations:
\begin{align}
\label{z14:x300} \varepsilon\int_0^\infty \xi
\left(\frac{1}{e^{b(\xi+\kappa)}-1}-
\frac{N}{e^{bN(\xi+\kappa)}-1}\right)d\sqrt{\xi} &= \cal E,
\\
\label{z15:x300} \int_0^\infty\left(\frac{1}{e^{b(\xi+\kappa)}-1}-
\frac{N}{e^{bN(\xi+\kappa)}-1}\right) d\sqrt{\xi}&= N.
\end{align}

 First, let us determine~$b$
from the relation
\begin{equation}\label{b:x300}
n=\frac 12\int_0^\infty\frac{\sqrt{\xi} \,d\xi}{e^{b\xi}-1}
=\frac{1}{b^{3/2}}\int_0^\infty\frac{\sqrt{\xi}
\,d\xi}{e^{\xi}-1}\,.
\end{equation}

 Hence
\begin{equation}\label{bb1:x300}
b=\frac{1}{(2n)^{2/3}\left(\int^\infty_0 \frac{\sqrt{\xi}
\,d\xi}{e^\xi-1}\right)^{2/3}}\,.
\end{equation}

 We use the following identity:
\begin{equation}\label{bb3:x300}
\int^\infty_0\left(\frac{1}{x^2}- \frac{1}{x^2(1+\alpha
x^2)}\right)\,dx=\sqrt{\alpha}\int_0^\infty\frac{\,d\xi}{1+\xi^2}\,.
\end{equation}

 Let us find
$N_{cr}$ from the relation
\begin{equation}\label{bb4:x300}
 N_{cr} = \frac12\int \left(\frac{1}{\sqrt{\xi}(e^{b\xi}-1)} -
\frac{N_{cr}}{\sqrt{\xi}e^{bN_{cr}\xi}-1}\right) \,d\xi =
\frac{1}{\sqrt{b}} \int_0^\infty\left(\frac{1}{e^{\xi^2}-1}-
\frac{N_{cr}}{e^{\xi^2N_{cr}}-1}\right) \,d\xi.
\end{equation}

 Subtracting
${1}/{\xi^2}$ from both terms of the difference, we obtain
$k_{0}=N_{cr}=k_0$, which determines the transition to the Bose
condensate,
\begin{align}
k_0&=\frac{1}{\sqrt{b}}\int\left(\frac{1}{e^{\xi^2}-1}
-\frac{1}{\xi^2}\right) + \frac{1}{\sqrt{b}}
\int\left(\frac{1}{\xi^2}-\frac{1}{\xi^2(1+\frac{k_{0}}{2}\xi^2)}
\right)\,d\xi \nonumber \\&\qquad -\frac{1}{\sqrt{b}} \int \left(
\frac{k_{0}}{e^{k_{0}\xi^2}-1} -
\frac{k_{0}}{k_{0}\xi^2(1+\frac{k_{0}}{2}\xi^2)}\right)\,d\xi.
\end{align}

 In view of the relations
\begin{equation}\label{b4a:x300}
\frac{1}{\sqrt{b}} \int \left(
\frac{k_{0}}{e^{k_{0}\xi^2}-1}-\frac{k_{0}}{k_{0}\xi^2(1+\frac{k_{0}}{2}\xi^2)}\right)\,d\xi
=\frac{\sqrt{k_{0}}}{\sqrt{b}} \int_0^\infty \left(
\frac{1}{e^{\eta^2}-1} -\frac{1}{\eta^2(1+\frac{\eta^2}{2})}
\right)d\eta
\end{equation}
and
\begin{equation}\label{b5:x300}
\frac{1}{\eta^2(1+\frac{\eta^2}{2})} = \frac{1}{\eta^2} -\frac
12\frac{1}{(1+\frac{\eta^2}{2})}
\end{equation}
solving the quadratic equation for $x=\sqrt{k_{0}}$, we obtain
$k_0\approx 4c^2n^{2/3}$, where
$$
c=\int_0^\infty\left(\frac{1}{e^{\xi^2}-1}-\frac{1}{\xi^2}\right)\,d\xi
\left(\frac12\int_0^\infty\frac{\sqrt{\xi}\,d\xi}{e^\xi-1}\right)^{2/3},
$$

\begin{lemma}\label{lemma3:x300}
 Let
$n> k\gg k_0$.
 Suppose that
$\kappa=-\mu$ and $\mu>0$.
 In this case, Eqs.~\eqref{z14:x300} and \eqref{z15:x300} have the solutions
$k^{-1/3}\gg\mu > k^{-1/3-\delta}$, where $\delta>\nobreak 0$ is
arbitrarily small.

 Consider relations~\eqref{z14:x300} and \eqref{z15:x300}.
 Making the change of variables
$\xi-\mu =\eta$ and then $b\eta=\varphi$, we obtain
\begin{align}
n&=\frac{1}{b^{3/2}}\int_0^\mu \frac{k\xi
d\sqrt{\xi}}{1-e^{-k\xi}}- \frac{1}{b^{3/2}}\int_0^\mu \frac{\xi
d\sqrt{\xi}}{1-e^{-\xi}}+
\frac{1}{b^{3/2}}\int_0^\infty\left(\frac{\xi}{e^\xi-1}-\frac{k\xi}{e^{k\xi}-1}\right)
d\sqrt{\xi}, \label{z10:x300}
\\ k&=\frac{1}{\sqrt{b}}\int_\varepsilon^\mu
\left(\frac{k}{1-e^{-k\xi}}-\frac{1}{1-e^{-\xi}}\right)d\sqrt{\xi}
+\frac 1{\sqrt{b}}\int_\varepsilon^\infty\left(\frac{1}{e^\xi-1} -
\frac{k}{e^{\xi k}-1}\right)d\sqrt{\xi}. \label{z11:x300}
\end{align}

 After making the change
$k\xi=x$ in the corresponding integrals, we see that, as
$k\to\infty$,
\begin{align}
n&=\frac{1}{b^{3/2}} \frac{1}{\sqrt{k}} \int_0^{\mu k} \frac{x
d\sqrt{x}}{1-e^{-x}}+o\left(\frac{1}{b^{3/2}}\right), \label{z12:x300} \\
k&=\frac{1}{\sqrt{b}} \int_\mu^{\mu k}
\frac{d\sqrt{x}}{1-e^{-x}}+o\left(\frac{\ln k}{\sqrt{b}}\right).
\label{z13:x300}
\end{align}
 Hence
$(\mu k)^{3/2} \ll \sqrt{k}$; in particular, $k\gg(\mu k)^3 >
k^{1-\delta}$ for any $1>\delta>0$.
\end{lemma}
 The estimate of the entropy is similar to that in Lemma~\ref{lemma2:x300}.

\begin{theorem}\label{theorem2:x300}
{1)} Suppose that $N_i$ is the number of particles on the
level~$\varepsilon i$, $N\leq k_0$, and all the variants
$\{N_i\}$ satisfying the condition
$$
\text{ $\sum_{i=1}^\infty {N_i}=N$, }
$$
are equiprobable under the condition
$$
\text{ $\varepsilon\sum_{i=0}^\infty |i|^{2} N_i=\cal{E}$. }
$$
 Then the difference
\begin{equation}\label{21a:x300}
b^{1/2} \sum_{i=1}^\infty N_i \varphi(bi)-\frac 12\int_0^\infty
\varphi(x)\frac{1}{|x|^{1/2}}\left\{ \frac{1}{e^{x+\kappa}-1}
-\frac{N}{e^{N(x+\kappa)}-1}\right\} \,dx,
\end{equation}
where the function $\varphi(x)$ is any bounded piecewise smooth
function continuous at the points~$bi$, tends in probability to
zero as $N\to\infty$.

{2)} Now let $N> k_0+o(k_0)$.

 The ratio of the number of variants in which the number~$N_0$
of particles in the condensate occurs more rarely than the number
\begin{equation}\label{th4:x300}
 N-4c^2n^{2/3} -o(n^{2/3}),
\end{equation}
where
$$
c=\int_0^\infty\left(\frac{1}{e^{\xi^2}-1}-\frac{1}{\xi^2}\right)\,d\xi
\left(\frac12\int_0^\infty\frac{\sqrt{\xi}\,d\xi}{e^\xi-1}\right)^{2/3},
$$
to the total number of variants tends to zero faster than
$$
e^{-(N-4c^2n^{2/3})^{1/3-\delta}},
$$
where $\delta$, $1/2>\delta>0$, is any arbitrarily small number
independent of~$N_0$.
\end{theorem}

 Now consider the case of the dimension
$0<d<2$.

What does it mean from from the viewpoint of Koroviev's conjuring
trick? Suppose that tickets were sold for different prices. The
most expensive ticket provides its owner with the largest
area~$S_{\max}$ in the auditorium and, therefore, the largest
area for collecting the bank notes. The prices of other tickets
decrease monotonically with the size of the area alotted to
spectators. If all the areas are the same and equal
to~$S_{\max}$, as in Koroviev's trick, then we have the situation
considered in Theorem~1, which corresponds to dimension $d=2$.
But if the areas decrease in size, then the dimension is $d<2$.
Let us find the threshold numbers in this situation.

 Let us find constants
$b$ and~$\kappa$ from the following relations:
\begin{align}
\int_0^\infty \xi\left\{\frac{1}{e^{b(\xi+\kappa)}-1}
-\frac{k}{e^{bk(\xi+\kappa-1)} }\right\}d\xi^\alpha &= n,
\\
\int_0^\infty\left\{\frac{1}{e^{b(\xi+\kappa)}-1}
-\frac{k}{e^{bk(\xi+\kappa-1)} }\right\}d\xi^\alpha &= k.
\end{align}
 For
$\kappa=0$, we have
\begin{equation}\label{ff1:x300}
n= \int \frac{\xi \,d\xi^\alpha}{e^{b\xi}-1} =
\frac{1}{b^{1+\alpha}}\int_0^\infty\frac{\eta
d\eta^\alpha}{e^\eta-1}.
\end{equation}
 Hence
\begin{equation}\label{ff2:x300}
b=\frac{1}{n^{1/(1+\alpha)}}\left(\int_0^\infty\frac{\xi
\,d\xi^\alpha}{e^\xi-1}\right)^{1/(1+\alpha)}.
\end{equation}
 Then, assuming the probabilistic limit to be equal to~$n$,
for the critical number~$k_0$, we obtain
\begin{align}\label{ff3:x300}
k_0&=\int_0^\infty
\left\{\frac{1}{e^{b\xi}-1}-\frac{k_0}{e^{k_0b\eta}-1}
\right\}\,d\xi^\alpha
\nonumber \\
&=\frac{1}{b^\alpha}\int_0^\infty\left(\frac{1}{e^\xi-1}
-\frac{1}{\xi}\right)\,d\xi^\alpha
+\frac{1}{b^\alpha}\int_0^\infty \left( \frac{1}{\xi}-
\frac{1}{\xi(1+(k_0/2)\xi)}\right)\,d\xi^\alpha \nonumber
\\&\qquad - \frac{ k_0^{1-\alpha}}{b^\alpha}\int_0^\infty\left\{
\frac{k_0^\alpha}{e^{k_0\xi}-1}
-\frac{k_0^\alpha}{k_0\xi(1+(k_0/2)\xi)} \right\}\,d\xi^\alpha.
\end{align}

 Denote
$$
c=\int_0^\infty \left(\frac 1 \xi -\frac{1}{e^{\xi}-1}\right)
\,d\xi^{\alpha}.
$$

 After the replacement
$k_0\xi=\eta$, we obtain
\begin{align}
\label{ff4:x300} &\frac{
k_0^{1-\alpha}}{b^\alpha}\int_0^\infty\left\{
\frac{k_0^\alpha}{e^{\eta}-1} -\frac{k_0^\alpha}{\eta(1+\eta/2)}
\right\}\,d\xi^\alpha
\nonumber \\
&\qquad= \frac{k_0^{1-\alpha}}{b^\alpha}\int_0^\infty
\left\{\frac{1}{e^\eta-1}-\frac{1}{\eta(1+\eta/2)}\right\}d\eta^\alpha
\nonumber \\
&\qquad=
\frac{k_0^{1-\alpha}}{b^\alpha}\left\{\int_0^\infty\left(\frac{1}{e^\eta-1}-\frac{1}{\eta}\right)+
\int_0^\infty\frac{d\eta^\alpha}{2(1+ \frac \eta 2)}\right\}= -c
\frac{k_0^{1-\alpha}}{b^\alpha}+
c_1\frac{k_0^{1-\alpha}}{b^\alpha}\,.
\end{align}
 Since
$$
\text{ $\frac{1}{\eta(1+\eta/2)}=\frac
{1}{\eta}-\frac{1}{2(1+\eta/2)}$\,, }
$$
denoting
$$
c_1= \int_0^\infty\frac{d\eta^\alpha}{2(1+\frac \eta 2)}\,,
$$
we can write
\begin{equation}\label{f5:x300}
\int_0^{\infty}\left(\frac{1}{\xi}
-\frac{1}{\xi(1+\frac{k_0}{2}\xi)}\right) \,d\xi^\alpha
=\frac{k_0}{2} \int_0^\infty
\frac{\,d\xi^\alpha}{1+\frac{k_0}{2}\xi}=
\left(\frac{k_0}{2}\right)^{1-\alpha}\int_0^{\infty}
\frac{d\eta^\alpha}{1+\eta}=c_1\left(\frac{k_0}{2}\right)^{1-\alpha}.
\end{equation}
 Hence
\begin{align}\label{f6:x300}
k_0
&=-\frac{1}{b^\alpha}c_1+\frac{1}{b^\alpha}c\left(\frac{k_0}{2}\right)^{1-\alpha}
- \frac{k_0^{1-\alpha}}{b^\alpha} \int_0^\infty
\left\{\frac{1}{e^\eta-1}-\frac{1}{\eta(1-\frac \eta 2)
}\right\}d\eta^\alpha
-\frac 12 \int\frac{d\eta^\alpha}{1+\frac\eta 2} \cdot
\frac{k_0^{1-\alpha}}{b^\alpha}
\nonumber \\
&=-\frac{1}{b^\alpha}c+\frac{k_0^{1-\alpha}}{b^\alpha}c.
\end{align}

 Therefore,
\begin{equation}\label{fff6:x300}
k_0 =k_0(\alpha) \approx c^{1/\alpha}n^{1/(1+\alpha)}\left(
\int^\infty_0 \frac{\xi
\,d\xi^\alpha}{e^\xi-1}\right)^{-1/(1+\alpha)}+o\left(n^{1/(1+\alpha)}\right).
\end{equation}

\begin{lemma}\label{lemma4:x300}
 Let
$n> k\gg k_0$.
 Suppose that
$\kappa=-\mu$ and $\mu>0$.
 In this case, Eqs.~\eqref{z14:x300} and \eqref{z15:x300} have the solutions
$$
\text{ $k^{-1/(\alpha+1)} \gg\mu > k^{-1/(\alpha+1)-\delta}$, }
$$
where $\delta>\nobreak 0$ is arbitrarily small.

 Consider the relations~\eqref{z14:x300} and \eqref{z15:x300}.
 Making the change of variables
$\xi-\mu =\eta$ and then $b\eta=\varphi$, we obtain
\begin{align}
n&=\frac{1}{b^{\alpha+1}}\int_0^\mu \frac{k\xi
\,d\xi^{\alpha}}{1-e^{-k\xi}}- \frac{1}{b^{\alpha+1}}\int_0^\mu
\frac{\xi \,d\xi^{\alpha}}{1-e^{-\xi}}+
\frac{1}{b^{\alpha+1}}\int_0^\infty\left(\frac{\xi}{e^\xi-1}-\frac{k\xi}{e^{k\xi}-1}\right)
\,d\xi^\alpha, \label{zz10:x300}
\\ k&= \frac
1{b^\alpha}\int_\varepsilon^\mu
\left(\frac{k}{1-e^{-k\xi}}-\frac{1}{1-e^{-\xi}}\right)\,d\xi^\alpha
+\frac 1{b^\alpha}\int_\varepsilon^\infty\left(\frac{1}{e^\xi-1} -
\frac{k}{e^{\xi k}-1}\right)\,d\xi^{\alpha}. \label{zz11:x300}
\end{align}

 After making the change
$k\xi =x$ in the corresponding integrals, we see that, as
$k\to\infty$,
\begin{align}
n&=\frac{1}{b^{\alpha+1}} \frac{1}{k^\alpha} \int_0^{\mu k}
\frac{x
d{x^\alpha}}{1-e^{-x}}+o\left(\frac{1}{b^2}\right), \label{zz12:x300} \\
k&=\frac{1}{b^\alpha} \int_\mu^{\mu k}
\frac{\,dx^\alpha}{1-e^{-x}}+o\left(\frac{\ln k}{b^\alpha}\right).
\label{zz13:x300}
\end{align}
 Hence
$(\mu k)^{\alpha+1} \ll k^{\alpha}$; in particular,
$$
\text{ $k^{\alpha/(\alpha+1)}\gg\mu k >
k^{\alpha/(\alpha+1)-\delta}$, }
$$
for any $1>\delta>0$.
\end{lemma}
\begin{equation}\label{ff6:x300}
k_0 =k_0(\alpha) \approx c^{1/\alpha}n^{1/(1+\alpha)}\left(
\int^\infty_0 \frac{\xi
\,d\xi^\alpha}{e^\xi-1}\right)^{-1/(1+\alpha)}+o\left(n^{1/(1+\alpha)}\right).
\end{equation}
 The estimate of the entropy is similar to that given in Lemma~\ref{lemma2:x300}.

 Thus, we have proved the following theorem.

\begin{theorem}
\label{theorem3:x300} {1)} Let the dimension be $0<d<2$, let
$\alpha=d/2$, and let
$$
\widetilde k_0(\alpha) = c^{1/\alpha}n^{1/(1+\alpha)}\bigg(
\int^\infty_0 \frac{\xi
\,d\xi^\alpha}{e^\xi-1}\bigg)^{-1/(1+\alpha)}\!\!,
\quad\text{where}\quad c=\int_0^\infty \bigg(\frac 1 \xi
-\frac{1}{e^{\xi}-1}\bigg) \,d\xi^{\alpha}.
$$

Let all the variants
$\{N_i\}$ satisfying the condition
$$
\text{ $\sum_{i=0}^\infty N_i=k$, }
$$
are equiprobable under the condition
$$
\text{ $\sum_{i=0}^\infty i^{1/\alpha}N_i=n$, }
$$
where $n$ is not integer, and let $k<\widetilde k_0(\alpha)$.

 Then the difference
\begin{equation}\label{21b:x300}
b^{\alpha} \sum_{i=1}^\infty N_i \varphi(bi)-\alpha\int_0^\infty
\frac{\varphi(x)}{x^{1-\alpha}}\left\{ \frac{1}{e^{x+\kappa}-1}
-\frac{k}{e^{k(x+\kappa)}-1}\right\} \,dx,
\end{equation}
where the function $\varphi(x)$ is any bounded piecewise smooth
function continuous at the points~$b_i$, tends in probability to
zero as $k\to\infty$.

{2)} Let $k> \widetilde k_0(\alpha) + o(\widetilde k_0(\alpha))$.
Then
\begin{equation}
{\mathbf P} \Big\{\big| N_0-\big[k-\widetilde
k_0(\alpha)\big]\big|\geq \delta_1 n^{1/(1+\alpha)} \Big\} \leq
e^{-|k-\widetilde k_0(\alpha)|^{\alpha/(\alpha+1) - \delta}},
\label{53a}
\end{equation}
where $\delta$ and $\delta_1$ are as small as desired and
independent of~$n$ and $\mathbf P$ stands for the ratio of the
number of variants satisfying the condition in the brackets to
the total number of versions corresponding to the value
$k=\widetilde k_0(\alpha)$.

The elements of the sequence $N_1,N_2,\dots,N_i$, $i\leq A/b$,
where $A$ is an arbitrarily large number independent of~$b$,
satisfy the relation
$$
b^{\alpha} \sum_{i=1}^{[A/b]} N_i \varphi(bi)-\int_0^\infty
\frac{\varphi(x)}{x^{1-\alpha}}\left\{ \frac{1}{e^{x}-1}
-\alpha\frac{\widetilde k_0(\alpha)}{e^{\widetilde
k_0(\alpha)x}-1}\right\} \,dx \xrightarrow{\mathsf P} 0
\qquad\text{as}\quad n\to\infty.
$$
\end{theorem}

In conclusion, let us show how the theorems given above can be
applied to debts.

 Consider the decreasing sequence~$l_i$ of durations, and let
$l_1$ denote the largest duration (the largest debt repayment
time). Suppose that the weak limit of the sum of the debts with
respect to a certain time interval corresponding to this duration
is $\overline{N}_i$.\footnote{If the durations $l_i$ and
$l_{i+1}$ vary by jumps, then we must fill in the intervals by
virtual durations, debts, and flows so that the sum of the debts
and flows would not change in the final result.} Consider the
sequence of \textit{flows} ${\cal{E}}_i= \overline{N}_i l_1/l_i$.

 Let
${\cal{E}}_0= \overline{N}_{i_0} l_1/l_{i_0}$ be the minimal
value of this series for all~$i$. Let us extrapolate
${\cal{E}}_i/{\cal{E}}_0$,\, $i=1, \dots$\,, by a smooth
curve~$\lambda(x)$.

Suppose that the integrals
$$
I_1(b)= \int_0^\infty \frac{dx}{e^{b\lambda(x)}-1} \qquad
\text{and} \qquad I_2(b)= \int_0^\infty
\frac{\lambda(x)}{e^{b\lambda(x)}-1}\,dx
$$
are convergent. Then the threshold value with respect to the mean
duration $T=\sum {\cal{E}}_i/\sum \overline{N}_i$ is of the form
${I_2(b)}/{I_1(b)}$, i.e., we must have $T\ge {I_2(b)}/{I_1(b)}$;
otherwise, \textit{a debt crisis occurs}
(see~\cite{Arxiv_Gas:x300},~\cite{TMFGibbs:x300}).

 The parameter~$b$ can be determined as follows. Consider the sum
$ M_s= \sum_{i=s}^{i=as} \overline{N}_il_1/l_i $ for sufficiently
large $s\sim 1/b$, i.e., for flows with sufficient short
durations on the interval $x=sb$, $x=asb$, such that $\lambda(x)$
is the line segment $\lambda'(x) =k$. Then, in view
of~\cite{QuantEconomics:x300,TMFGibbs:x300},
$$
M_s\cong \int_{bs}^{as}\frac{kx\
  dx}{e^{kx}-1}, \qquad
\sum_{i=s}^{as} \overline{N}_i \cong \frac 1b\int_{bs}^{abs}
\frac{dx}{(e^{kx}-1)}.
$$
 Hence the relation
$$
T_s=\sum_{i=s}^{a s} \frac{\overline{N}_i
l_1}{l_i}\bigg/\sum_{i=s}^{a s} \overline{N}_i \sim \frac 1b
\int_{bs}^{a bs} \frac{kx\ dx}{e^{kx}-1}\bigg/\int_{bs}^{a bs}
\frac{dx}{(e^{kx}-1)}
$$
yields the asymptotic values of~$b$.

In the foregoing, we proved theorems on threshold values in the
most important case where the integral~$I_1$ is divergent. How
can we apply them to the debt crisis problem.

If the relation $2>d>0$ holds asymptotically on a segment near
$\lambda(x)=0$, then, in view of Theorem~3, the threshold value
of the relative mean duration~$T$ is $ T_{\text{cr}}
=c_0c^{-1/\alpha} b^{-\alpha} $, where the quantity~$b$ is
calculated as described above and $c_0=\int_0^\infty
\xi(e^\xi-1)^{-1}\,d\xi^\alpha$.

In particular cases, the introduction of a multicurrency system
(an increase in the share of barter) can be calculated by
combining Bose statistics with Boltzmann statistics.

 Let us note the following.
 To change
$N_1$ into $N_0$, we must shift the numbering by one.
 Then the number
$n$ will change.
 It
is easy to see that, instead of~$n$, one must take $\overline{n}$
equal to $n-k$
(compare with formulas (2)--(5)).

 Laws of economics and physics have much in common,
which was noted by Adam Smith as early as the 18th century.
 At present, there is a special branch of science called econophysics.
 On the other hand, economic problems helped the author
to better understand the mechanism of some physical
laws~\cite{QuantEconomics:x300}, namely, the nature of Bose
condensate and describe it in deterministic terms.

 There is also the so-called impulse (explosive) flicker noise.
 There were no explanations or formulas for this well-known phenomenon.
 The author provided a mathematical explanation for this fact
also related to the parastatistics given above.

 The ``explosive flicker noise'' of the Chernobyl atomic power
station posed~\cite{Chernobyl:x300}, as it were, a problem for
the author, who had become an expert in passages to the limit.
 Judging by the good
agreement with experiments (see, for
example,~\cite{RJMP_15_1:x300}, \cite{Noy:x300},
\cite{Joseph:x300}, \cite{Exper:x300}), this problem had been
solved before the ``anniversary'' of this tragic event.

\newpage

\appendix
\section{General Addenda}

{\vskip-3pt}\rightline{\sl``\dots it is a terrible thing for a
man to find out suddenly}

\rightline{\sl that all his life he has been speaking nothing but
the truth.''}

\rightline{\bf Oscar Wilde, ``The Importance of Being Earnest,''
1895.}
\medskip

In the years 1989--1991, the author spent a great deal of time
trying  explain to the country's  leading economists and
politicians the relevant mathematical results concerning the
coming catastrophe in the USSR economy.

 {\bf The author's articles in the Soviet press  (1989--1991)}

The situation at that time, when the cost of one personal
computer was about the same as that of 1000 cubic meters of
timber, was reminiscent of the beginning of commerce in tropical
Africa, when European traders would exchange worthless little
mirrors for ebony and gold. Having in mind economists and
politicians (not mathematicians), I called the relevant
mathematics  ``tropical,'' \footnote{ See the author's article in
the journal {\textit Kommunist}, No.~13, 1989. }\ since I felt
that the  term ``idempotent analysis" used by mathematicians for
this branch of mathematics  would sound disharmonious to the ear
of an economist or a politician.

Although at that time I was able to publish articles in
newspapers and even talk to some of the highest political
leaders, I did not succeed in convincing anyone.  Before the 1998
default, I also tried, but not as hard.

Here is a list of my publications in the Soviet press in
1989-1991. \vskip-1pt

1. ``Are we destined to foretell?", {\textit{Kommunist}}, No.~13,
1989, pp.~89--91.\vskip-1pt

2. ``Who has what prices?", {\textit Pravda}, March 8,
1991.\vskip-1pt

3. ``The ruble under the dollar's thumb (how to restore the blood
flow in our economy)," {\textit {Nezavisimaya Gazeta}}, March 26,
1991.\vskip-1pt

4. ``And all the goods will flow abroad," {\textit {Torgovaya
Gazeta}}, No.~68, June 6, 1991.\vskip-1pt

5. ``The law of the broken thermos bottle," {\textit {Poisk}},
No.~3 (89), March 11, 1991.\vskip-1pt

6. ``How to avoid a total catastrophe," {\textit {Izvestiya}},
No.~187, August 7,1991. [On August 19, 1991, the putsch began,
and on December 8, 1991, The Belovezhskoe agreement on the
disintegration of the Soviet Union was signed.] \vskip-1pt

7. ``How does the state intend to pay its debts?", {\textit
{Torgovaya Gazeta}}, March 22, 1995.\vskip-1pt

8. ``How does a pound of rubles fare against one of their pound
sterling?", {\textit {Literaturnaya Gazeta}}, November 27, 1996.
\vskip-1pt

9. ``The authors of the catastrophe are those who strive to
become the saviors," {\textit {Izvestiya}}, September 24,
1998.\vskip-1pt

In the paper \cite{Masl_MZ_85_1:x300},  I touched upon  the
question of how hard it was  to explain to our leadership the
mathematical laws which indicated the way out of the crisis
cycle. It is even more difficult, I think, to explain them to
foreign politicians.

Let me quote an excerpt from an interview  I gave to the {\textit
Minnesota Daily\/} (published June 1, 1990, page 11, during
Gorbachev's visit to the US). ``Victor Maslov
$\langle\cdots\rangle$ said the biggest problem is that the
current Soviet leaders have no idea about the average citizen's
struggles. $\langle\cdots\rangle$ He said a large portion of the
real economy in the Soviet Union is dependent on the black market
-- a fact that goes unrecognized by the government. ``The
government has lots of good ideas but when they try to put these
ideas to our people, it is not good because they have not
investigated the real situation", Maslov said.
$\langle\cdots\rangle$ Maslov said it might be beneficial for
Soviet leaders to speak with economists here, but added that the
economists might be unfamiliar with the unique problems facing
the Soviet Union."

I had in mind the Nobel Prize winner V.~V.~Leontiev and Leo
Hurwicz (the future Nobel Prize winner), whom I succeeded in
convincing in the inexpediency of measures proposed to alleviate
the ``tropical'' situation in the USSR. But they both said that
they cannot intervene because they are not experts in the Soviet
economy.

{\bf Excerpta from ``Expertise and Experiments" ({\textit Novy
Mir}, no.~1, 1991)}

As an addition to what was explained in \cite{Masl_MZ_85_1:x300},
let me present some excerpts from my article in the once popular
journal {\textit Novy Mir}. The article was submitted to the
journal in 1989, but since  {\textit Novy Mir\/} did not appear at
all in 1990 (because of financial difficulties), the article was
published only in January 1991. In it, I developed the
conclusions that follow from the application of idempotent
analysis to economic problems of the USSR.

Since it is impossible to avoid mathematical rules, a second
currency, the dollar, began functioning in the country, the USSR
disintegrated, and the Islamic revolution spread to several
countries. The Pol Pot regime in Cambodia was crushed by my
father-in-law, Le Duan [Le Duc Anh],\footnote{See V.~Maslov,
``Daring to touch Radha", Lviv, Academic Express, 1993,
pp.~42--43; available at the website {\tt
http://www.viktor-maslov.narod.ru} in the Literature section; see
the discussion concerning the Pol Pot regime on pp.~27--28.}\ but
as an attractor (see below), it did not disappear.

\subsection{EXPERTISE AND EXPERIMENTS}

{\textit Excerpts from the author's article in  ``Novy Mir'',
No.~1, 1991.} The heading and comments in square brackets
$[\,\cdot\,]$ were added by the editor of the translation.
\vskip1pt

[{\bf Computer-aided panel of experts}]

It is natural [to save the Soviet economy] not to act by trial and
error, but to begin by  picking a panel of experts, representing
different sections of the population, to search for compromise
solutions, to analyze a huge number of variants, and choose the
optimal one among them.~$\langle\cdots\rangle$ It~is precisely a
panel of experts, expressing the averaged opinion of various
strata of society, their possible reaction, the means  used by
various sections of the population to circumvent certain laws and
regulations, that can replace  statistical data  and the study of
the rules, partially market-oriented, partially semi-legal, which
people employ in practice. $\langle\cdots\rangle$

The expert system must be conductive to an optimal and careful
intervention in the existing systems of social relationships,
taking into account the psychology of the Soviet
person.~$\langle\cdots\rangle$ It~is only the interaction of a
panel of experts with a computer that can succeed  in
establishing a stable equilibrium between the centralized, the
regional, and the market segments of the economy in their complex
mutual relations. The expert system allows testing out various
versions of rules and regulations not on the living population,
as had so unfortunately been done during the partial prohibition
of vodka, but by means of computer modeling aided by the panel of
experts. The result of such testing must be made public and be
verifiable.

If by means of effective measures we will succeed in stopping the
USSR economy from falling apart, which may lead to the actual
disintegration of the Soviet Union, then according to a simple
calculation, the optimal solution would not be integration with
Europe, but with Japan, South Korea, with countries where the
development of electronics in many years ahead of ours and far in
front, say, of that of heavy industry. \vskip1pt

[{\bf Three currencies}]

As an example, let me describe the solution of the problem of
making the ruble  convertible proposed by the expert system.
$\langle\cdots\rangle$

The impossibility of spending rubles to buy needed goods and the
instability   of the situation in the country, as well as the
instinctive equalitarian attitude of various segments of the
population, impelled people who had amassed large sums in rubles
to exchange them for dollars and deposit the dollars in foreign
banks. These circumstances lead to the depreciation of Soviet
made goods, their flow out of the country, and thus to colossal
economic losses for the state. $\langle\cdots\rangle$

The presently existing [in 1989]  huge discrepancy in the exchange
rate of the ruble against the dollar for different types of goods
and services  contributes to the deepening of the economic
crisis, and, as computer modeling has shown, it cannot be
overcome in the framework of the two-currency system  without a
significant  increase of social tension. $\langle\cdots\rangle$

We are familiar with the three-currency system introduced during
the NEP period [the New Economic Policy in Soviet Russia in the
1920ies] involving the ruble, the chervonets [banknote equivalent
to the 10 ruble gold coin], and the dollar, as well as the
three-currency system presently working in different regions of
Russia and involving the ruble, the dollar, and the ``certificate
for specific goods'' (the latter can be exchanged for the type of
merchandise specified on it).

The expert panel proposed a rather paradoxical compromise version:
rubles, dollars, ``general-purpose certificates.'' The latter are
coupons or cards that can be exchanged for a wide choice of goods.
This version is the one that won the contest supervised by
computers together with the panel of experts.  Thus the system
most manageable by means of price regulation is described by  the
following model.

The salaries in rubles remain the same, but in addition to the
salary, employees are given all-purpose certificates of nominal
value equal to the salary for the same position in the 1980ies.
Using these certificates, one would be able to purchase essential
goods for rubles, as well as certain  items needed for comfort
(automobiles, vacuum cleaners, TV sets, refrigerators). Besides,
these certificates or coupons could be used jointly with rubles
for services, rent, travel expenses. $\langle\cdots\rangle$

The fundamental question of what goods can be sold for dollars, or
for certificates plus rubles, and what priorities should be chosen
cannot be answered without the help of the panel of experts (and
ratified by the Supreme Soviet). $\langle\cdots\rangle$

The buying and selling of certificates on the black market should
not be forbidden or hampered.

The advantage of the proposed system is its flexibility and
stability, the weakness of its reaction to unexpected changes such
as strikes, popular uprisings, etc. The latter can only lead to a
higher inflation rate of the ruble, which will only accelerate the
passage to a convertible currency.

[{\bf Long-term forecast and attractors}]

The difference between an expert system giving the optimal recipe
at any given moment and a system providing a long-term recipe
(taking into account the far away future), is similar to the
difference between tactics and strategy. One can be a good
tactician but a poor strategist. The mathematical part of the
system that I have described is tactical. Its implementation is
laborious, but possible. But when it is necessary to give
long-term predictions while influencing their evolution, the
problem becomes significantly more complex.

The situation may be compared with the diagnosis and the therapy
of an illness. The diagnostic expert system, which assists the
group of doctors in choosing the best therapy for the patient at
the given moment, crucially differs from the expert system
predicting the course of the illness in the long term while
influencing its development. The patient has several options: to
be totally cured, or becoming a chronic sufferer of the disease,
and so on. Any of the options that may be realized in the long
term will be called an attractor. Clearly, in order to create
such an expert system, it is necessary to input into the
computer's memory all the possible pathways for evolving to the
chronic state. Besides the optimal trajectories of the changes in
the patient's health resulting from the prescribed therapy, the
computer must show which of these trajectories lead to one of the
various attractors. $\langle\cdots\rangle$\vskip-1pt

A similar, but much more complicated mathematical problem arises
in creating an expert system with strategic goals that would model
social and economic phenomena. Such a system must take into
account social, political, and economic regimes that existed in
the past, and those that arose recently. These can be the regimes
implemented in Western countries, the national-socialist systems
of Spain, Italy, or Germany, and, finally, ``unique'' regimes
(the Khomeny regime in Iran, the ``communist'' regime of Pol Pot
in Cambodia), which have no analogs, but are theoretically
possible. Their symptoms must be studied and input in the
computer's memory. If the trajectories of solutions tend towards
such attractors, this must be taken in consideration.
$\langle\cdots\rangle$\vskip-1pt

Such a strange attractor arises in the situation around Germany,
where East German citizens either emigrated to West Germany in
order to earn hard currency there, or didn't work at all, since
their salaries in the West would be 10 times larger. They were
awaiting reunification with West Germany as a kind of emigration
together with their homes and factories. The computer
extrapolates this into the following ``strange'' attractor: at
first, East Germany will rejoin West Germany, later so will
Poland (under less favorable conditions), then the Baltic states,
then the Russian Federation. The result will be an economic
conquest of the territories that Hitler
dreamed~of.~$\langle\cdots\rangle$\vskip-1pt

In studying long-term processes, we essentially lose the
possibility of efficiently using the experts, since they will
have to describe the evolution of the world outlook of the
corresponding social stratum in twenty-thirty years time on the
basis of their present reaction. Nevertheless, the experts should
still be called upon to predict the changes in the basic
viewpoint of the given segment of society.
$\langle\cdots\rangle$\vskip-1pt

If for the initial axiom we accept the existence of evolution laws
not yet understood, it is just as absurd to complain about them
and condemn their functions as it would be to accuse Archimedes'
principle of the responsibility of somebody drowning. The laws
must be studied and one must act within a sufficiently wide
framework of these laws. $\langle\cdots\rangle$\vskip-1pt

\nopagebreak

This would be a first step towards the understanding of the actual
laws of human society. It is extremely difficult to give a
mathematical formalization  of long-term evolution (unlike the
case of a tactical expert system), nevertheless we should strive
to such an understanding. At least in order to give a forecast
from the outside.

\section{Addenda for the  physicists}
\subsection{Phase transition to the condensate state
and from the condensate state as transition to another type of
energy}

The principle of ``social-economic profitability'' exemplified by
the grouping of firms (spectators) into cells was carried over by
the author to the Bose gas condensate: the same grouping into
dimers, trimers, and clusters occurs in an ideal gas according to
the author's conception.
 The ideal gas is a model in which
particles are regarded as points, and hence there exists only
translational motion.
 If the temperature~$T$
drops below $T_{\text{cr}}$, then the number $N-N_{\text{cr}}$ of
particles crosses into the Bose condensate with large
probability, and this, according to the author's
theory~\cite{Arxiv_Gas:x300} means that points combine into pairs
(dimers); therefore, we now have the vibratory motion of
oscillators (three-dimensional oscillators or six-dimensional
particles), which implies that the heat capacity is proportional
to~$T^3$ (see the figure\footnote{This graph is taken from the
textbook\cite{Kvasn:x300} and approximately describes the
situation with the real Bose gas Helium-4. For the ideal limiting
case, the experiment was first explained by the author
in~\cite{TMF_2009:x300}.}).
 If the
gas is rotated at a certain rate or flows in a capillary with a
given velocity or convection of the ideal gas is assumed, then
some pairs acquire additional integrals of motion and we obtain a
mixture of two-dimensional and six-dimensional particles.
 This
explains the Thiess--Landau two-liquid model.
 In addition, as is
seen from the calculation given below, two-dimensional particles
(the one-dimensional oscillator) at a lower temperature
$T_\lambda<T_0$ furnish a heat capacity of the form of the
$\lambda$--point near the critical value~$T_\lambda$ (see
Theorem~1).
 Therefore, this concept is consistent with the
Thiess--Landau model and with the concept of $\lambda$-point
\cite{TMF_2009:x300}.

\begin{figure}
\includegraphics{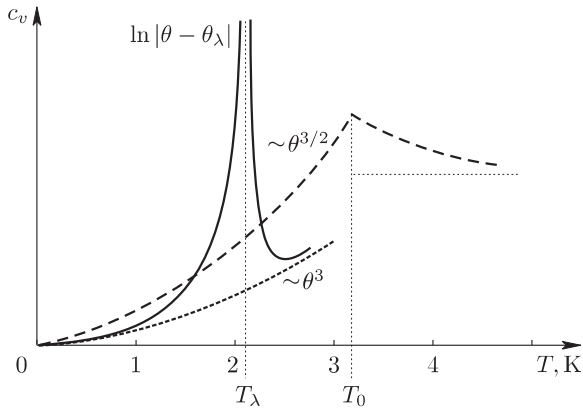}
\caption{The graph of the heat capacity $c_v$ for a quantum Bose
gas. In the region of temperatures $0<\theta<1$ on the Kelvin
scale, the dependence $c_v \sim \theta^3$ is typical, and so is
the dependence $-\ln|\theta_{\lambda} - \theta|$ in the region of
the $\lambda$-point.
 The graph of the heat capacity of an ideal Bose gas
is depicted by the dotted line.} \label{fig1}
\end{figure}

 The energy
$\cal{E}$ defined at the beginning of Sec.~3 is called global.
 The quantity
$\cal{E} \gg\varepsilon$ is a constant nonstandard number.
 We studied the
$N$-particle Schr\"odinger equation for symmetric gas
\textit{molecules} interacting in a constant volume~$V$ by the
Lennard-Jones law (see~\cite{Masl_RJ_14_3:x300}).

 It has turned out that, in the limit, as
$N$ (density) increases, the interaction simultaneously decreases,
while the energy $\cal{E}\gg\varepsilon$ remains constant, there
occurs (under ultrasecond quantization~\cite{Book_Ultra:x300}
generalizing the Bardeen--Cooper--Schrieffer method for Cooper
pairs) a phase transition for $N>N_{\text{cr}}$ from one spectral
series~\cite{Quasi_Part:x300} to two other series associated with
superfluidity (one-dimensional vibrational motion $d=2$ and
three-dimensional chaotic vibrational motion $d=6$).

 In essence,
the particular form of the Lennard-Jones potential in these limit
transitions was of no importance; we used only the fact of the
existence of a discrete negative (hole) spectrum contributing to
the formation of vibrating pairs~\cite{Masl_RJ_15_4:x300}.

 Transitions of such type are useful in economic theory,
because they are general and are independent of the form of
interaction.
 The transition to the spectrum corresponding to superfluidity
is one from chaotic state to joint motion (convection of a gas).
 A similar situation occurs for flicker noises.

 Here the phase transition explains the passage from
the individualistic desire to get rich quickly (in Koroviev's
trick, the situation for $k<\sqrt{n}$) to the collective
unification with one common aim in mind (convection; in the
conjuring trick, the situation for $k\gg \sqrt{n} \ln\ n$). The
phase transition to a condensate shows that the cluster
individualization of dimers is partially preserved.

 In our example of the transition to the condensate state
of an ideal Bose gas, there occurs a partial unification into
dimers (clusterization) and partial superfluidity.
 It is the latter phenomenon that leads to the phase transition
of zeroth order (the author invoked this transition to explain
the so-called spouting effect of Allen and
Jones)~\cite{Masl_Arxiv_Econ:x300}.

 Recall that such transitions from one type of energy
to other types occur in local spectral series for a fixed
constant, but very large global energy~$\cal{E}$.

 We have regarded the transition to the Bose-condensate state as
transition of the kinetic energy of an ideal three-dimensional
gas to the vibrational energy of a two-particle ideal (two-point
chaotic) gas.
 The appearance of the
$\lambda$-point was interpreted by the author as partial
convection, and hence, in view of the momentum conservation law,
as one-dimensional vibrational (two-dimensional chaotic)
dimension~\cite{RJ_2009:x300}.
 This explains the two-liquid Thiess--Landau model.
A detailed justification of this interpretation was given
in~\cite{RJ_16_2:x300}.

Turbulence was interpreted by the author as transition from the
condensate state of a nonchaotic laminary flow to the state of
partially chaotic turbulent flow and, therefore, as transition
from one type of energy to another
type~\cite{Masl_TMF_1993:x300},~\cite{Masl_Shafar:x300}.

 The Boltzmann equation bewildered mathematicians and
physicists by its disagreement with the laws of the classical
mechanics of many particles.
 We explain this discrepancy by the
phase transition of the classical condensate (nonchaotic) state
to the condensate chaotic one.
 Only using this approach,
can one obtain a rigorous deduction of the Boltzmann equation.
 The
transition to the condensate state yields the solution of the
Gibbs paradox problem~\cite{Arxiv_Gas:x300}.

 One of the important transitions of explosive noise energy is the transition
from the potential energy of overhanging snow or rock avalanches
to the kinetic energy of moving avalanches, which depends on the
random values of the mass and height of the avalanche.
 The
critical value at which this transition can occur, just as that
of explosive flicker noise, can be calculated by means of the
spectral expansion of the time series.

 It is known that
both the sociological
statistics~\cite{Masl_Social:x300},~\cite{Masl_Whoiswho:x300}
and animal population statistics correspond to flicker noises.

\subsection{On Explosive Flicker Noises}

Consider the time series
$$
b_0,b_1, b_2, \dots, b_s, \ \ \sum^s_{i=0} b_i^2  \gg 1.
$$

Using Kotelnikov's theorem, we obtain the Fourier series
$$
f(t) = \sum_{i=0}^s a_i \cos\biggl(\frac{\pi it}{s}\biggr)
$$
assuming that the values $b_j$,\, $j=0,1, \dots, s$, are  given
at integer points $t=j$.
 Therefore,

1) $\frac 1s\sum_{i=0}^s a_i^2 = \frac 1s\sum^s_{i=0}
b_i^2=A_0$,\, $A_0 s\gg 1 $;

2) the energy~$E_s$ is
$$
E_s= \frac 1{A_0}\int_0^{2\pi}\biggl(\frac{df}{dt}\biggr)^2 (t)\,
dt.
$$
 Hence we have
\begin{equation}\label{0tmf:y900}
\frac 1{A_0}\sum_{i=0}^s i^2 a^2_i =\frac{E_s s^2}{\pi^2}.
\end{equation}

 As is well known, by the \textit{spectral density} of an interval $l < s$
we mean~\cite{Vovk:y900} the mean square of the amplitude; namely,
$$
A_l = \frac{1}{s-l} \sum_{i=l}^s a^2_i.
$$

 These definitions can be applied to the case of ordinary noises
if take into account their stochastic nature; see, for
example,~\cite{Vovk:y900}.

 We have white noise when the spectral density does not decrease.
 We have a Wiener process when the spectral density
decreases as $l^{-2}$,\, $s\gg l$.

 Here we define explosive flicker noise as noise in the sense
of the paper~\cite{Masl_85_3:y900} in Secs.~2 and~3.

 Let $\cal{E}$ denote the total energy
\begin{equation}\label{e1:y900}
{\cal E} =\frac{1}{A_0}\sum_{i=0}^s i^2a_i^2= \frac{E_s
s^2}{\pi^2}, \quad s\gg 1.
\end{equation}

\begin{definition}
 By explosive flicker noise we mean a sequence
$\{a_i^2 s_i\}^s_{i=1}$, where the $s_i\geq 0$ are integers, such
that all the collections $\{a_i s_i\}^s_{i=1}$ are equiprobable
under the conditions
\begin{equation}\label{e3:y900}
\sum_{i=0}^s s_i \leq s, \quad \left[\frac{1}{A_0}\sum_{i=0}^s
a^2_i s_i\right] =s, \quad \frac{1}{A_0}\sum_{i=0}^s i^2a_i^2 s_i
={\cal E}.
\end{equation}
\end{definition}

 The quantity~${\cal E}$ is called the \textit{global energy of explosive
flicker noise} and $E_s$ is its temperature. The process of
transition from~$s$ to~$s+1$ is called an indeterminate process
in a sense contrary to the sense of Markov.

Thus, the probability of occurrence of any sequence from the
collection $\{a_i^2 s_i\}^s_{i=1}$ coincides with the probability
of occurrence  of the sequence $a_i^2, \ i=0,1, 2, \dots, s$
satisfying conditions~\eqref{e3:y900}. Such a
collection $\{a_i^2 s_i\}$ is said to be
\textit{trajectory-probability equivalent} to the sequence
$a_i^2$,\, $i=0,1,2, \dots, s$.

This means that the input sequence ${a_i^2}/{A_0}$ is a random
sample from the given equiprobable collection. This sample is
``smooth'' in contrast to the other variants of the collection,
which ``oscillates'' about this sample. Only it satisfies the
condition on the spectral density (as though it is the weight of
the trajectory~$s_ia_i^2$,\, $i=0,1,2, \dots, s$) Therefore,
although we consider ``spectral density'', these concepts do not
coincide with universally accepted ones, and this fact can lead
to a misunderstanding among specialists in the field of random
processes. But here we use these concepts in another sense and for
special types of noises, namely, explosive flicker noises.

 For explosive flicker noises, we present a theorem similar
to Theorem~3 from~\cite{Masl_85_3:y900}. But, first, let us find
the threshold value for~$E_s$ as a function of~$s$. We have
\begin{equation}
\frac{\cal E}{s}=\frac{E_s s}{\pi^2}\,.
\end{equation}
The threshold value for this ratio was, in fact, calculated in
Theorems 1-3 from~\cite{Masl_85_3:y900}, because using the
notation  $s=N$ in Theorem~2 in that paper or that in Theorems~1
and~3, we obtain ${\cal E}=n$,\, $s=k$.

In~\cite{Masl_85_3:y900}, the critical value $k_0$ was calculated
as a function of~$n$, and if $k_0$ is given just as in the above
case, then the ratio $n/k$ as a function of~$k$ yields the
critical value $E_s^{\text{cr}}$ as a function of~$s$. This is
similar to the calculation of the critical temperature for a Bose
gas using $N_{\text{cr}}$~\cite{Landau:y900}. Then Theorem~3
from~\cite{Masl_85_3:y900}, where $\alpha$ is \textit{the rate of
decrease of the spectral density}, can be carried over to
explosive flicker noises in the following form.

 Let us define the constants $\beta$ and~$\kappa$ from the relations
\begin{align}\label{e4:y900}
\int_0^\infty \xi \left(\frac{1}{e^{\beta(\xi+\kappa)}-1}-
\frac{s}{e^{\beta s(\xi+\kappa)}-1}\right)d\xi^\gamma &= {\cal E},
\\
\label{e5:y900}
\int_0^\infty\left(\frac{1}{e^{\beta(\xi+\kappa)}-1}-
\frac{s}{e^{\beta s(\xi+\kappa)}-1}\right) d\xi^\gamma&= s,
\end{align}
where $\gamma=1/2-\alpha/4>0$.

Denote
$$
c= \int_0^\infty \left(\frac 1\xi-\frac{1}{e^\xi-1}\right)
d\xi^{\gamma},  \qquad c_0= \int_0^\infty \frac{\xi
d\xi^\gamma}{e^\xi-1}\,.
$$
Then
\begin{equation}\label{e6:y900}
E_s^{\text{cr}}(\alpha)=\pi^2 c_0c^{-1/\gamma}\beta^{-\gamma},
\end{equation}
where $\beta$ is defined from  \eqref{e4:y900} for~$\kappa=0$.

\begin{theorem}
{1)} Let $E_s\geq E_s^{\text{cr}}(\alpha)$. Let $0<\alpha <2$,
and let all the variants
 $\{s_i a_i^2\}$
satisfying the condition
$$
\left[\frac{1}{A_0}\sum_{i=0}^s a^2_i s_i\right]=s
$$
be equiprobable under the condition
$$
\frac 1A_0\sum_{i=0}^\infty |i]^{2}a^2_i s_i={\cal E}.
$$
Then the difference
\begin{equation}\label{e7:y900}
\frac{\beta^{\gamma}}{A_0} \sum_{i=1}^s a^2_i s_i \varphi(\beta
i)-\gamma\int_0^\infty \varphi(x)\frac{1}{|x|^{1-\gamma}}\left\{
\frac{1}{e^{x+\kappa}-1} -\frac{s}{e^{s(x+\kappa)}-1}\right\} dx,
\end{equation}
where the function $\varphi (x)$ is any bounded piecewise smooth
(with finitely many discontinuities) function continuous at the
points $\beta i$,
tends in probability to zero as $s\to\infty$.

{2)} Let $E_s < E_s^{\text{cr}}(\alpha)$. Then, for $s_0$, the
following ratio holds:
\begin{equation}\label{e8:y900}
{\bold P} \Big\{\big|s_0-\big[s-\widetilde
s(\alpha)\big]\big|\geq \delta_1\widetilde s(\alpha)\Big\} \leq
e^{-|s-\widetilde s(\alpha)|^{\gamma/(1+\gamma)}},
\end{equation}
where
$$
\widetilde s(\alpha)=c^{1/\gamma} c_0^{1/(1+\gamma)} {\cal
E}^{1/(1+\gamma)}
$$
and where $\delta$ and $\delta_1$ are as small as desired and
independent of~${\cal E}$ and $\mathbf P$ stands for the ratio of
the number of variants satisfying the condition in the brackets
to the total number of versions corresponding to the value
$s=\widetilde s(\alpha)$.
 For $i<\widetilde s(\alpha) +o(\widetilde s(\alpha))$,
$i< A/\beta$, where $A$ is an arbitrarily large number
independent of~$\beta$, the following relation holds:
$$
\frac{\beta^{\gamma}}{A_0} \sum_{i=1}^{[A/\beta]} a_i^2 s_i
\varphi(\beta i)-\gamma\int_0^\infty
\varphi(x)\frac{1}{|x|^{1-\gamma}}\left\{ \frac{1}{e^{x}-1}
-\frac{\widetilde s(\alpha)}{e^{\widetilde s(\alpha)x}-1}\right\}
dx \xrightarrow{\mathsf P} 0 \qquad\text{as}\quad s\to\infty.
$$
\end{theorem}

 In view of the relation
$\sum_{i=o}^s s_i<s$, this means that $s_0$ equal to $s-
\widetilde s(\alpha)$ occurs in many variants with probability
$$
\text{ $1-\exp\{-|s-\widetilde s(\alpha)|^{\gamma/(1+\gamma) -
\delta}\}$. }
$$

For $\gamma =0$,
we have an analog of Theorem~1 from~\cite{Masl_85_3:y900}.

 The second part of the theorem means an explosive transition to another
kind of energy for $s\gg s_0(\alpha)$ if the global energy ${\cal
E}$ remaind constant, because $s_0a_0^2$ makes no contribution
into the global energy of explosive flicker noise
(cf.~\cite{Koverda:y900}).

 This can explain both the transition of turbulent flow
(flicker noise according to the Taylor conjecture for
Kolmogorov's law) to laminar flow and the transition of an
irreversible process described by the Boltzmann equation and
Boltzmann's $H$-theorem to a fully reversible deterministic
process with vanishing entropy governed by the laws of the
classical mechanics of many particles.

\begin{figure}[h]
\includegraphics[angle=90,width=\textwidth]{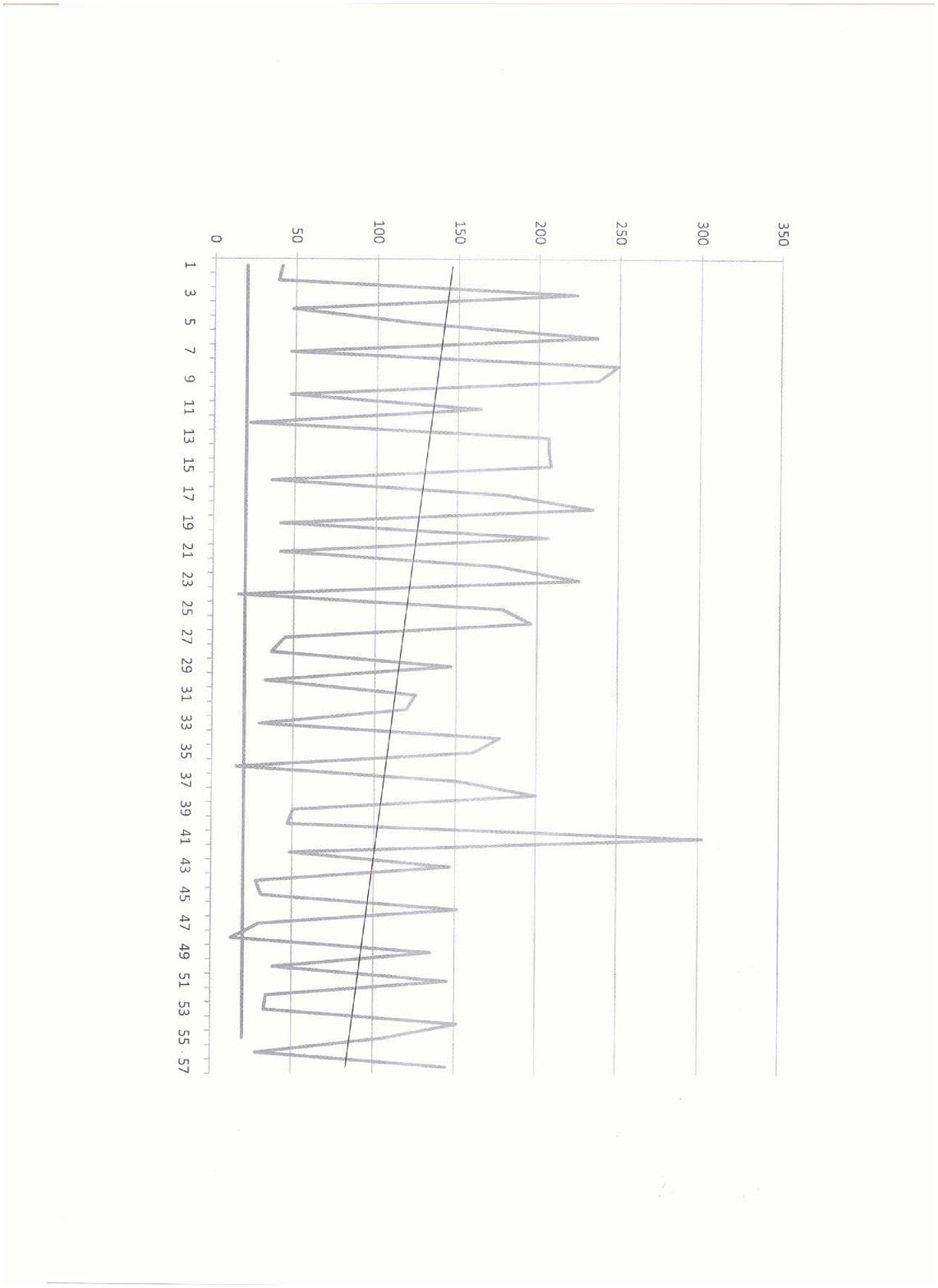}
\caption{An example of the average money flow.
Abscissa axis: days; ordinate axis: average} \label{fig2}
\end{figure}

Eighty years ago, the phenomenon
of frequency-proportional fluctuations in electric voltage,
i.e., the flickering noise~\cite{Kogan},
was discovered in electronic devices.
In processes with such a spectrum,
the energy is pumped from high-frequency to low-frequency fluctuations,
and there is a possibility of large-scale catastrophic outbursts.
I would not solve this mysterious problem,
just as the mysterious problem of the turbulence origination,
if I did not notice that the amplitudes of oscillations
about the average value of debt flows were tremendous
(see Fig.~2).
Both problems are in fact equivalent to each other
in their mathematical statement.

The transition to turbulence is a phase transition from the Taylor
spectrum \cite{Kogan}--\cite{masl-om-94-2} to the Kolmogorov spectrum.

\end{document}